\documentstyle[12pt]{article}
\begin{document}

% ***************    NEW COMMANDS   *******************
\newcommand{\bcc}{\begin{center}}
\newcommand{\ecc}{\end{center}}
\def \inbar{\vrule height1.5ex width.4pt depth0pt}
\def \C{\relax\hbox{\kern.25em$\inbar\kern-.3em{\rm C}$}}
\def \R{\relax{\rm I\kern-.18em R}}
\newcommand{\Z}{\ Z \hspace{-.08in}Z}
\newcommand{\be}{\begin{equation}}
\newcommand{\ee}{\end{equation}}
\newcommand{\bea}{\begin{eqnarray}}
\newcommand{\eea}{\end{eqnarray}}
\newcommand{\p}{\psi}
\renewcommand{\l}{\epsilon}
\newcommand{\f}{\phi}
\newcommand{\g}{\gamma}
\newcommand{\G}{\Gamma}
\newcommand{\e}{\eta}
\newcommand{\z}{\pi}
\newcommand{\m}{\mu}
\newcommand{\n}{\nu}
\newcommand{\s}{\sigma}
\renewcommand{\t}{\tau}
\renewcommand{\a}{\alpha}
\renewcommand{\b}{\beta}
\newcommand{\k}{\kappa}
\renewcommand{\d}{\delta}
\newcommand{\r}{\rho}
\newcommand{\th}{\theta}
\renewcommand{\pt}{\frac{\partial}{\partial t}}
\newcommand{\ppt}{\frac{\partial^{2}}{\partial t^{2}}}
\newcommand{\nn}{\nonumber}
\renewcommand{\ll}{\left[ }
\newcommand{\rr}{\right] }
\newcommand{\kt}{\rangle}
\newcommand{\br}{\langle}
\newcommand{\fs}{\small}
\newcommand{\so}{S_{0}}
\newcommand{\I}{\mbox{1}_{m\times m}}
\newcommand{\In}{\mbox{1}_{n\times n}}
\newcommand{\xo}{x_{0}}
\newcommand{\po}{\psi_{0}}
\newcommand{\eo}{\e_{0}}
\newcommand{\ts}{\tilde{S}}
\newcommand{\pss}{\frac{\partial}{\partial s}}
\newcommand{\tcuf}{\tilde{\cal F}}
\newcommand{\cuf}{{\cal F}}
\newcommand{\oot}{\mbox{\fs$\frac{1}{2}$}}
\newcommand{\iot}{\mbox{\fs$\frac{i}{2}$}}
\newcommand{\cur}{{\cal R}}
\newcommand{\iv}{\imath\! v}
\newcommand{\ib}{\int_{0}^{\b}}
\newcommand{\lll}{\left( }
\newcommand{\rrr}{\right)}
\newcommand{\llc}{\left\{ }
\newcommand{\rrc}{\right\} }
\newcommand{\lpt}{\left.}
\newcommand{\rpt}{\right.}
\newcommand{\rar}{\longrightarrow}
\newcommand{\lar}{\longleftarrow}
\newcommand{\cinf}{C^{\infty}\!}
\newcommand{\vol}{d\mbox{\bf\fs$\Omega$}}
\newcommand{\sx}{\mbox{\fs$(x)$}}
\newcommand{\hD}{\hat{D}}
\newcommand{\V}{(V)}
\newcommand{\La}{\Lambda}
\newcommand{\ph}{{\cal P}({\cal H})}
\newcommand{\cp}{\C\! P}
\renewcommand{\l}{\lambda}
\title{Geometric Phase, Bundle Classification, and Group Representation}
\author{Ali Mostafazadeh\\ \\ Center for Relativity \\
Department of Physics \\ The University of Texas at Austin \\
Austin, Texas 78712 }
\maketitle
\baselineskip=18pt
\begin{abstract}
The line bundles which arise in the
holonomy interpretations of the geometric phase display curious similarities
to those encountered in the statement of the Borel-Weil-Bott theorem of the
representation theory. The remarkable relation of the geometric phase to
the classification of complex line bundles provides the necessary
tools for establishing the relevance of the Borel-Weil-Bott theorem to
Berry's adiabatic phase. This enables one to define a set of
topological charges for arbitrary compact connected semisimple
dynamical Lie groups.
In this paper, the problem of the determination of the parameter space
of the Hamiltonian is also addressed. A simple topological argument is
presented to indicate the relation between the Riemannian structure
on the parameter space and  Berry's connection.
The results about the fibre bundles
and group theory are used to introduce a procedure to
reduce the problem of the  non-adiabatic (geometric) phase
to Berry's  adiabatic phase for cranked Hamiltonians. Finally,
the possible relevance of the topological charges of the
geometric phase to those of the non-abelian monopoles is pointed
out.
\end{abstract}
\section{Introduction}
In the past ten years, since the revival  of the geometric phase,
\cite{pal,mead}, by Berry \cite{berry}, the subject has attracted
the attention of many physicists. The main reason  for the unusual
popularity of this remarkably simple subject, particularly among
the theoretical physicists,  has been its rich mathematical and
physical foundations.

Recently, it was shown that the two holonomy interpretations
of Berry's phase were linked via the theory of universal
bundles, \cite{bbmr,mb}. This remarkable coincidence of the physics
of  geometric phase and the mathematics of fibre bundles enables
one to set up a convenient framework to analyze the non-adiabatic
phase \cite{mb}. In the present paper, the results of \cite{mb} are
briefly reviewed and their generalization to arbitrary finite dimensional
unitary systems are presented.

In section 2, it is shown how the study of the standard example of
a spin in a precessing magnetic field directs one to the
Borel-Weil-Bott (BWB) theorem of the representation theory of compact
semisimple Lie groups. In section 3, the relation of BWB theorem
to the phenomenon of geometric phase is discussed in a general
setting. Section 4 is devoted to a discussion of the relation
of Berry's connection and the Riemannian geometry of the
parameter space. Section 5 includes the discussion of the reduction
of the non-adiabatic phase problem to the adiabatic one
for the cranked Hamiltonians.
Section 6 consists of a short account on the classification
of the parameter spaces and the topology of non-abelian monopoles.
Section 7 includes the conclusions.
A short proof of a result of Floquet theory is presented in the
appendix.
\section{Bundle Classification and the Holonomy Interpretations
of the Geometric Phase}
There are two mathematical interpretations of Berry's (adiabatic)
phase. These are due to Simon \cite{simon}, and Aharonov and
Anandan \cite{aa}. I shall refer to these two approaches by
``BS'' and ``AA'' which are the abbreviations of ``Berry-Simon''
and  ``Aharonov-Anandan'', respectively.

In the BS approach, one constructs a line bundle $L$ over the space
$M$ of the parameters of the system. Then, $L$ is endowed with a
particular connection which reproduces Berry's phase as the
holonomy of the closed loop in the parameter space.

Let us consider a quantum mechanical system whose evolution is
governed by a parameter dependent Hamiltonian:
\[ H=H(x) \;\;\;\; ,\;\;\;\; x\in M\; .\]
Assume that for all $x\in M$ the spectrum of $H(x)$ is discrete and that
there are no level crossings. Then, locally one can choose a
set of orthonormal basic eigenstate vectors $\{ |n,x\kt\}$.
As functions of $x$, $|n,x\kt$ are smooth and  single valued. By definition,
they satisfy:
\be
H(x)|n,x\kt =E_{n}(x)|n,x\kt \; ,
\label{q1}
\ee
where $E_{n}(x)$ are the corresponding energy eigenvalues.
The Hamiltonian is made expilicitly time dependent by interpreting
time $t$ as the parameter of a curve
\be
C: [0,T]\ni t\rar x(t)\in M \; ,
\label{q1.3}
\ee
and setting
\be
H(t):=H(x(t))\;\;\;\; ,\;\;\;\; t\in [0,T]\; .
\label{q1.5}
\ee
Then, each closed curve $C$ in $M$ defines a periodic Hamiltonian
with period $T$. I shall discuss only the evolution of nondegenerate
cyclic states with period $T$.

Under the adiabatic approximation the initial eigenstates undergo
cyclic evolutions, \cite{berry}. If $|\p_{n}(t)\kt$ denotes the
evolving state vector, i.e., the solution of the Schr\"{o}dinger
equation:
\bea
H(t)|\p_{n}(t)\kt& =& i\frac{d}{dt}|\p_{n}(t)\kt
\label{q3} \\
|\p_{n}(0)\kt&:=&|n,x(0)\kt \; , \nn
\eea
then
\be
|\p_{n}(T)\kt\br\p_{n}(T)|\simeq |\p_{n}(0)\kt \br\p_{n}(0)|\; .
\label{q2}
\ee
After a cycle is completed, the state vector gains a phase factor which
consists of a dynamical ($e^{i\omega}$) and a geometric ($e^{i\g}$)
part
\be
|\p_{n}(T)\kt =e^{i(\omega+\g )}|\p_{n}(0)\kt\; ,
\label{q4}
\ee
where
\[ \omega :=-\int_{0}^{T}E_{n}(x(t))\, dt \; ,\]
and
\bea
 e^{i\g}&:=& \exp \oint_{C} A \label{q5} \\
A&:=&-\br n,x|d|n,x\kt = -\br n,x|\frac{\partial}{\partial x^{\m}}|n,x\kt
\, dx^{\m} \; . \label{q6}
\eea
The one-form $A$ is known as  Berry's connection one-form,\cite{berry}.

In \cite{simon}, Simon showed that $A$ could be interpreted as a connection
one-form on a (spectral) line bundle $L$ over $M$,
\be
\C\rar L\rar M\; ,
\label{q7}
\ee
whose fibres are given by the energy eigenrays in the Hilbert space ${\cal H}$
\be
L_{x}:=\{ z|n,x\kt : z\in \C\}\; .
\label{q8}
\ee
Thus, in the BS approach Berry's phase is identified
with the holonomy of the loop $C\subset M$ defined by the
connection one-form $A$ of eq.\ (\ref{q6}).

In the AA approach one considers a complex line bundle $E$, or
alternatively the associated $U(1)$-principal bundle, over the projective
Hilbert space $P({\cal H})=\C\! P^{N}$, $N:=dim({\cal H})-1$ :
	\be
 	\C\rar E \rar\ph\; .
	\label{q9}
	\ee
The fibres are the rays, i.e., $\forall \e=|\e\kt\br\e|\in\ph$
	\be
	E_{\e}:=\{z|\e\kt : z\in \C\} \; .
	\label{q10}
	\ee
The AA connection one-form ${\cal A}$ \cite{aa} is then viewed as a connection
one-form on $E$ and the geometric phase is identified with the corresponding
holonomy of loops
	\be
	{\cal C}: [0,T]\ni t\rar\e (t)\in\ph \; ,
	\label{q11}
	\ee
in $\ph$.
In the adiabatic approximation one approximates $\e (t)$ by
$\p_{n}(t)$ of eq.\ (\ref{q3}).

These two interpretations of Berry's phase turn out to be linked
via the theory of {\em universal bundles}. It is
shown in \cite{bbmr,mb} that $E$ (with $N\to\infty$)
is indeed the universal classifying
line bundle \cite{egh,cd2,nakahara}, and as a result of the
classification theorem
for complex line bundles \cite{cd2,egh,isham}, every complex line bundle
can be obtained as a pullback bundle from $E$. In particular, there is
a smooth map
	\be
	f: M\rar\ph
	\label{q12}
	\ee
such that
	\be
	L=f^{*}(E)\; .
	\label{q12.1}
	\ee
The map $f$ is simply given by
	\be
	f(x):=|n,x\kt\br n,x|\; .
	\label{q13}
	\ee
Furthermore, the fact that the phase is obtained from either of
$A$ or ${\cal A}$ is a consequence of the theory of {\em universal
connections} \cite{nr,sch}. In fact, the AA connection ${\cal A}$
is precisely the universal connection which yields all connections
on all complex line bundles as pullback connections. In particular,
Berry's connection on $L$ is given by
	\be
	A=f^{*}({\cal A})\; .
	\label{q14}
	\ee
These results are exploited in \cite{mb} to explore the quantum dynamics
of Berry's original example:
	\be
	H(x)=b\,\vec{x}.\vec{J}\;\;\;\; ,\;\;\;\;\vec{x}\in S^{2}\subset
        \R^{3}\; ,
	\label{q15}
	\ee
where $b$ is the Larmor frequency, $\vec{x}$ is the direction of the
magnetic field, and $\vec{J}=(J_{i}),\, i=1,2,3$,
are  the generators of rotations, $J_{i}\in so(3)=su(2)$.
In \cite{mb}, it is shown that if one considers the case of precessing
magnetic field, i.e., precessing $\vec{x}$, about a fixed axis then
one can promote  Simon's construction
to the non-adiabatic case, namely, define a non-adiabatic analog
of Berry's connection and identify the non-adiabatic phase  with
its holonomy. This can be done in general unless the frequency
of precession, $\omega$, becomes equal to $b$.
In the northern hemisphere the non-adiabatic connection
$\tilde{A}$ is given by
	\be
	\tilde{A}=ik(1-\cos \tilde{\theta} )\, d\phi
	\label{q16}\; ,
	\ee
where $k$ labels an eigenvalue of $H(x)$ (alternatively an eigenvalue
of $J_{3}$), and
	\bea
	\cos \tilde{\theta}&:=&\frac{\cos \theta -\nu}{\sqrt{
	\n^{2}-2\n\cos\theta +1}}\; ,\label{q17} \\
	\n &:=&\frac{\omega}{b} \; .
	\label{q18}
	\eea
Here $(\theta ,\phi )$ are the spherical coordinates ($\theta\in
[0,\pi )$), and $\nu$ is the ``slowness parameter,'' \cite{berry2}.
The adiabatic limit  is characterized by $\nu\rightarrow 0$. In this limit
$\tilde{A}$ approaches to Berry's connection
	\be
	A=ik(1-\cos\theta )\, d\phi\; .
	\label{q18.1}
	\ee
The topology of a line bundle on $S^{2}$ is determined by its first Chern
number
	\be
	c_{1}:=\frac{i}{2\pi}\int_{S^{2}}\Omega\; ,
	\label{q19}
	\ee
where $\Omega$ is the curvature two-form. For line bundles,
the curvature two-form is
obtained from the connection one-form by taking its ordinary exterior
derivative
\cite{cd1}. A simple calculation shows that taking $\Omega=d\tilde{A}$
results in
	\be
	c_{1}=-2k \;\;\;\;\mbox{for}\;\; \n <1\; .
	\label{q20}
	\ee
This is quite remarkable since the fact that $c_{1}$ is an integer
agrees with the fact that $k$ is a half-integer. The first statement
is an algebraic topological result, whereas the second is related to
group theory. One of the best known mathematical results that links
these two disciplines is the celebrated Borel-Weil-Bott (BWB) theorem
\cite{bott,wallach,fh,segal}.

Eq.\ (\ref{q20}) may also be viewed as
an example of a topological quantization of angular momentum. In the
language of magnetic monopoles, which are relevent to the adiabatic
case, $k=-c_{1}/2$ corresponds to the product of the electric and
magnetic charges \cite{goddard,monopole}.

\section{Borel-Weil-Bott Theorem and the Berry-Simon Line bundles}
The BWB theorem constructs all the  finite dimensional irreducible
representations
(irreps.) of semisimple compact Lie groups from the irreps. of their
maximal tori. The construction is as follows.

Let $G$ be a semisimple
compact Lie group and $T$ be a maximal torus. Let ${\cal G}$ and ${\Upsilon}$
be the Lie algebras of $G$ and $T$, respectively. $G$ can be viewed as a
principal bundle over the quotient space $G/T$, \cite{bb}:
	\be
	T\rar G\rar G/T \; .
	\label{q21}
	\ee
The homogeneous space $G/T$ can be shown to have a canonical complex
structure \cite{wallach}. Since $T$ is abelian, its irreps. are one
dimensional \cite{bb}. Thus, each irrep. $\Lambda$ of $T$ defines
an associated complex line bundle $L_{\Lambda}$ to  (\ref{q21}):
	\be
	\C\rar L_{\Lambda}\rar G/T\; .
	\label{q22}
	\ee
Now, consider a $\La$ whose corresponding line bundle
$L_{\La}$ is an ample (positive) line bundle. Then,
$L_{\La}$ has the structure of a holomorphic line bundle. BWB theorem
asserts that all the irreps. of $G$ are realized on the spaces of
holomorphic sections of ample (positive) line bundles,
$L_{\La}$. In particular, the space
${\cal H}_{\Lambda}$ of the holomorphic sections of $L_{\La}$
provides the irrep. of $G$ with maximal weight $\La$, \cite{fh,wallach,segal}.

The simplest nontrivial example of the application of BWB theorem
is for $G=SU(2)$. In this case, $T=U(1)=S^{1}$ and $G/T=S^{2}=\cp^{1}$.
The bundle (\ref{q21}) is the Hopf bundle, \cite{bb}:
	\be
	U(1)=S^{1}\rar SU(2)=S^{3}\rar S^{2} \; .
	\label{q22.1}
	\ee
$\La$ takes nonnegative half-integers. It is usually denoted by $j$
in QM. It is a common knowledge that $j=0,\oot,1,\cdots$ yield
all the irreps. of $SU(2)$ and that the $j$-representation has
dimension $2j+1$. The dimension of the space ${\cal H}_{\La}$
can be given by an index theorem \cite{fh,bott}. For $SU(2)$ it  is
obtained by the Riemann-Roch theorem in the context
of the theory of Riemann surfaces. The result is
	\be
	dim\,({\cal H}_{\La})=c(L_{\La})=1+c_{1}(L_{\La})\; ,
	\label{q23}
	\ee
where $c$ and $c_{1}$ denote the total and the first Chern
numbers of $L_{\La}$. This means that one must have:
	\be
	c_{1}(L_{\La})=2j\; .
	\label{24}
	\ee
Combinning (\ref{q20}) and (\ref{24}), one recovers the line
bundle $L_{\La}$ as Simon's line bundle $L$ of (\ref{q7}) for
$k=-j$.

In the rest of this section, I shall try to show that there is
a general relationship between the constructions used in the BWB theorem
and those encountered in BS interpretation of Berry's phase.
To proceed in this direction, let us consider the generalization
of (\ref{q15}) to an arbitrary compact semisimple Lie group,
namely consider:
	\be
	H(x)=\epsilon\sum_{i=1}^{d}x^{i}J_{i}\;\;\; ,\;\;\; (x^{i})
	\in \R^{d}-\{ 0\}\; .
	\label{q25}
	\ee
Here, $J_{i}$ are the generators of $G$, and $\epsilon$
is a constant with the dimension of energy. Since $H(x)$ is assumed
to be hermitian, $J_{i}$ must be represented by hermitian matrices.
In other words, the group $G$ is in a unitary representation.
In this sense, the example of $G=U(N)$ plays a universal
role.\footnote{This reminds one of the Peter-Weyl theorem, \cite{segal,bb}.}

The system described by eq.\ (\ref{q25}) is studied in \cite{as}
and \cite{wang}.
In \cite{as}, it is argued that in general there are unitary operators
$U(t)$ which diagonalize the instantaneous Hamiltonian:
	\be
	H(t)=U(t)\, H_{D}(t)\, U(t)^{\dagger} \; .
	\label{q26}
	\ee
In view of eq.\ (\ref{q1.5}), one has
	\be
	U(t)=U(x(t))\; ,
	\label{q27}
	\ee
where
	\be
	x(t)=\lll x^{i}(t)\rrr\in {\cal G}-\{ 0\} =\R^{d}-\{ 0\} \; ,
	\label{q28}
	\ee
are the points of the loop in the parameter space. In fact, one can
show that the parameter space ``is not'' $\R^{d}-\{ 0\}$ but a
submanifold of this space, namely the flag manifold $G/T$.

To see this, let me first introduce the root system of ${\cal G}$
associated with $\Upsilon$ and the corresponding Cartan
decomposition:
	\be
	{\cal G}_{\C}=\Upsilon_{\C}\oplus_{\a}{\cal G}_{\a}\; ,
	\label{q30}
	\ee
where the subscript $\C$ means {\em complexification}
and $\a $ stand for the roots.
Let $l$ denote the rank of ${\cal G}$, $\{ H_{i}\}_{i=1,2,\cdots ,l}$
and $E_{\a}$
be  bases of $\Upsilon$ and ${\cal G}_{\a}$, respectively
\cite{georgi,bb,fh,wallach}. Then, one has
	\bea
	\ll H_{i},H_{j} \rr &=&0\nn\\
	\ll H_{i},E_{\a}\rr & \propto & E_{\a} \nn \\
	\ll E_{\a},E_{-\a}\rr & \propto & H_{\a} \in \Upsilon \label{q31} \\
	\ll E_{\a},E_{\b}\rr &\propto & E_{\a +\b}\;\;\;\;\;\mbox{for}\;\;
	\b\neq -\a\; .\nn
	\eea

Any group element can be obtained as a product of the exponentials
of the generators of the algebra. In particular
	\be
	U(t)=\exp \ll i\sum_{\a}\chi_{\a}(t)\, E_{\a}\rr\,
	\exp \ll i\sum_{i}\chi_{i}(t)\, H_{i}\rr \; .
	\label{q32}
	\ee
Since any diagonal element commutes with $H_{i}$'s, it belongs to
$\Upsilon$. Hence, one has
	\be
	H_{D}(t)=\sum_{i}b_{i}(t)\, H_{i}\; .
	\label{q33}
	\ee
Substituting eq.\ (\ref{q33}) in eq.\ (\ref{q32}) and using the
resulting equation to simplify eq.\ (\ref{q26}), one obtains
	\bea
	H(t)&=&e^{i\sum_{\a}\chi_{\a}(t)\, E_{\a}}H_{D}(t)\,
	e^{-i\sum_{\a}\chi_{\a}(t)\, E_{\a}} \label{q34} \\
	&=&e^{i\sum_{\a >0}\ll z_{\a}(t)\, E_{\a}+z_{\a}^{*}(t)E_{-\a}\rr}
	H_{D}(t)\,e^{-i\sum_{\a >0}\ll z_{\a}^{*}(t)\, E_{\a}+z_{\a}(t)
	\, E_{-\a}\rr } \; . \label{q35}
	\eea
In eqs.\ (\ref{q34}) and (\ref{q35}) $\chi_{\a}\in \R$ and $z_{\a}\in\C$
are time dependent parameters. It is shown in \cite{as} that in general
the geometric phase is given in terms of $\chi_{\a}$'s or alternatively
in terms of $z_{\a}$'s, and it does not depend on $H_{D}(t)$. It is
not difficult to see that indeed $\chi_{\a}$  correspond to the
coordinates of the points of the flag manifold $G/T$. Alternatively,
one can use the complex coordinates $z_{\a}$. This is  reminiscent
of the fact that $G/T$ has a canonical complex structure, \cite{wallach}.
This completes the proof of the claim that the true parameter space
of the system described by (\ref{q25}) is $G/T$, or a submanifold
of $G/T$. I will come back to this point in section 6. The fact that
$G/T$ can be viewed as  embedded in ${\cal G}$ is useful because it
allows one to work with the global cartesian coordinate systems on
${\cal G}=\R^{d}$, \cite{wang}. A natural
embedding of $G/T$ is provided by taking a regular (non-degenerate)
element $H_{0}$ of $\Upsilon$
and considering the Adjoint action of $G$ on ${\cal G}$. The orbit
corresponding to $H_{0}$ is a copy of $G/T$. Thus, one might note that
in eq.\ (\ref{q25})
	\be
	x=(x^{i})\in G/T\subset \R^{d} \; .
	\label{q36}
	\ee

The fact that the phase information is encoded in $U(t)$ of eq.\ (\ref{q26})
can be used to simplify the problem, namely one can restrict to the
case where the $H_{D}(t)=H_{D}(0)=H_{0}$ is kept constant, i.e.,
	\be
	H_{D}=\sum_{i}b_{i}\, H_{i}=:H_{0}\in \Upsilon\;\;\;\; ,\;\;\;\; b_{i}
	={\rm const.} \; .
	\label{q37}
	\ee

The Hilbert space ${\cal H}$ of the quantum state vectors provides the
representation space. It can be decomposed into irrep. spaces.
I shall assume that ${\cal H}$ (or the subspace of ${\cal H}$
relevant to the geometric phase) corresponds to an irrep. with
maximal weight $\La$, \cite{fh}. The weights are the simultaneous
eigenvectors of $H_{i}$'s, \cite{georgi}. They are conveniently denoted
by $|\l_{1},\cdots ,\l_{l}\kt$, or collectively by $|\l\kt$, where
	\be
	H_{i} |\l\kt =\l_{i}|\l\kt\;, \;\;\;\forall i=1,\cdots l\; .
	\label{q38}
	\ee
Clearly, the weight vectors $|\l\kt$ are the eigenstate vectors of the
initial Hamiltonian. Here, I have set $U(0)=1$ in eq.\ (\ref{q26}), \cite{as}.
In general, this can be achieved by appropriately choosing the maximal torus
$T$. Thus, one has
	\be
	H(x(0))=H_{D}=H_{0}\; ,
	\label{q39}
	\ee
and
	\be
	H_{D}|\l\kt=\sum_{i=1}^{l}b_{i}\l_{i}\, |\l\kt \; .
	\label{q40}
	\ee
Making the dependence of $H_{D}$ ($H_{0}$) on the initial point $x_{0}:=
x(0)$ explicit, one can write eq.\ (\ref{q40}) in the form
	\bea
	H_{0}(\xo )|\l ,\xo\kt&=&E_{\l}(\xo )|\l ,\xo \kt \label{q41} \\
	E_{\l}(\xo )&:=&\sum_{i=1}^{l}b_{i}\, \l_{i}(\xo )\; . \nn
	\eea
The weight vectors $|\l ,\xo\kt$ are precisely the eigenvectors $|n,\xo\kt$
of the instantaneous Hamiltonian $H_{0}(\xo )$. Since $\xo$ can be
chosen arbitrarily, one can simply drop the subscript ``$0$'', i.e.,
replace $\xo$ by $x$ and $H_{0}(\xo)$ by $H(x)$.

The BS line bundle, in this case, is obtained as the pullback bundle
from the universal classifying bundle $E$,
	\be
	L_{\l}^{\rm BS}:=f^{*}(E)\; ,
	\label{q42}
	\ee
induced by the map
	\[ f:M\ni x\rar|\l ,x\kt\br\l ,x|\in\ph\subset\cp^\infty\; .\]
Recalling some basic facts about the flag manifolds and their relation
to projective spaces, \cite{fh}, one finds that in fact $L_{\l}^{\rm BS}$
corresponds to the line bundle $L_{\La}$ of the BWB theorem, if the weight
vector $|\l ,\xo\kt$ is chosen to be the maximal weight $\La$ of the
representation. First, let us recall, \cite{fh,wallach}, that flag manifolds
are projective varieties, i.e., there exist embeddings of $M$ into
$\cp^{\infty}$
	\be
	i:M\hookrightarrow\cp^{\infty}\; .
	\label{q44}
	\ee
Indeed, one can obtain $M=G/T$ as a unique closed orbit of the action of $G$
on ${\cal P}(\C^{N+1})=\cp^{N}$, for some $(N+1)$-dimensional irrep.,
\cite[\S 23.3]{fh}. The line bundle
$L_{\La}$ is then the restriction (pullback under the identity map)
of $E$:
	\be
	L_{\La}=i^{*}(E)\; .
	\label{q45}
	\ee

Let $|v_{0}\kt$ be a nonzero vector in the representation (Hilbert)
space  of the $\La$-representation of $G$, $G_{\C}$ be the complexification
of $G$ and consider the map
	\[ \Phi :G_{\C}\rar\ph \; ,\]
defined by
	\be
	\Phi (\tilde{g}):=\ll U(\tilde{g})|v_{0}\kt \rr=
	U(\tilde{g})|v_{0}\kt\br v_{0}|U(\tilde{g})^{\dagger}\; .
	\label{q46}
	\ee
Here $U(\tilde{g})$ is the representation of $\tilde{g}\in G_{\C}$ and
$\ll U(\tilde{g})|v_{0}\kt\rr$ denotes
the ray passing through $U(\tilde{g})|v_{0}\kt$. $\Phi$ is clearly
not one-to-one.
Let $P$ be the closed subgroup of $G_{\C}$ defined by
	\be
	P:=\left\{ \tilde{h}\in G_{\C}\: :\: U(\tilde{h})|v_{0}\kt=
	c|v_{0}\kt\: ,\:\mbox{for some}\; c\in\C -\{ 0\} \right\} \; .
	\label{q46.1}
	\ee
By construction the map $\Phi$ induces a one-to-one map on $G_{\C}/P$:
	\be
	\hat{\Phi}:G_{\C}/P\rar\ph \; .
	\label{q47}
	\ee
Now, let us choose
	\be
	|v_{0}\kt := |\La ,\xo\kt \; ,
	\label{q47.1}
	\ee
and denote by $B$ the {\em Borel subgroup} of $G_{\C}$ generated by
$H_{i}$ and $E_{\a >0}$.
Then, $B\subset P$ and consequently $G_{\C}/P$ is a compact submanifold
(subvariety) of $G_{\C}/B$. However, one has the identity
	\[ G_{\C}/B = G/T \:,\]
where by equality I mean the diffeomorphism of homogeneous spaces,
\cite{wallach}. Thus, in general $G_{\C}/P \subset G/T$.

The extreme case is when $P=B$, i.e., $M=G_{\C}/P=G/T$. However, in general
$B$ may be a proper subgroup of $P$, in which case the parameter
manifold can be restricted to the submanifold
$G_{\C}/P$ of $G/T$. This depends on the
representation, i.e., on $\La$.

Let us consider the general case, i.e., $M=G_{\C}/P$. The basic vectors
$|\l ,x\kt$ are parametrized by the points of $G_{\C}/P\subset G/T$
and the map $f$ of (\ref{q13}) becomes
	\be
	f:G_{\C}/P\ni x\rar|\l ,x\kt\br\l ,x|\in \ph\; .
	\label{q48}
	\ee
In view of the fact that $G_{\C}/P\subset G/T$,
one may work with the representative of $x=\ll g\rr\in G/T$ rather than
$x=\ll \tilde{g}\rr\in G_{\C}/P$ for the parameters $x$.
The next logical step is to compare the map $\hat{\Phi}$ of
(\ref{q47}) with $f$. Let $x\in M\subset G/T$, then every eigenstate vector
$|\l ,x\kt$ can be obtained by the action of $G$ on a nonzero vector.
In particular, there is a $g_{x}\in G$ such that
	\be
	|\l ,x\kt = U(g_{x})\, |\l ,\xo\kt \; .
	\label{q49}
	\ee
Combinning eqs.\ (\ref{q47.1}), (\ref{q48}), (\ref{q49}), and
specializing to $\l =\La$, one finds
	\be
	f(x)=U(g_{x})\, |v_{0}\kt\br v_{0}|\, U(g_{x})=\ll
	U(g_{x})|v_{0}\kt\rr \; .
	\label{q50}
	\ee
Recalling the procedure according
to which $x$ is assigned to represent the parameter, (\ref{q36}),
of the system (\ref{q25}), one can identify $\ll g_{x} \rr\in G_{\C}/P\subset
G/T$ with $x$, i.e.,
	\[ U(g_{x})\equiv U(x)\; ,\]
and consequently
	\be
	f(x)=\ll U(x)\, |v_{0}\kt\rr=\hat{\Phi}(x)\; .
	\label{q51}
	\ee
For the special case of $P=B$, the map $\hat{\Phi}$ becomes
the map $i$ of (\ref{q44}). Thus, according to eqs.\ (\ref{q45})
and (\ref{q51}) the following identity is established:
	\be
	L_{\La}=f^{*}(E)\; .
	\label{q52}
	\ee
Comparing eq.\ (\ref{q52}) with eq.\ (\ref{q42}), one arrives
at the desired result, namely that the bundle $L_{\La}$ of the
BWB theorem is identical to the $BS$ bundle $L_{\La}^{\rm BS}$.
In particular, the dimension of the irrep., i.e., the Hilbert
space ${\cal H}$ is given by the number of the linearly
independent holomorphic sections of $L_{\La}^{\rm BS}$. The latter
is a topological invariant of $L_{\La}^{\rm BS}$.

It is well-known that the topology of a complex line bundle is uniquely
determined by its first Chern class $\hat{c}_{1}$, \cite{milnor,mb}.
$\hat{c}_{1}$ is represented by a closed differential 2-form on $M$. It
can be characterized by a set of ($p:=dim\, H_{2}(M,\!\Z )$) integers
by integrating it over
$p$ compact 2-dimensional submanifolds of $M$ which are called the 2-cells
of $M$. For example, if $G=SU(2)$, $M=S^{2}$ and the space $S^{2}$ is the
only 2-cell. Therefore, $\hat{c}_{1}$ is determined by a single
integer $c_{1}$ via eq.\ (\ref{q19}).

In general, the following modification of eq.\ (\ref{q19}) provides
the necessary integers
	\be
	c_{1}^{a}=\hat{c}_{1}(\s_{a}):=\frac{i}{2\pi}\int_{\s_{a}}\Omega
	\; ,
	\label{q53}
	\ee
where $\s_{a}$ is the $a$-th 2-cell ($a=1,\cdots ,p$), $c_{1}^{a}$ is
the first Chern number associated with $\s_{a}$, and $\Omega$ is the
curvature 2-form of the line bundle.

For the case of the BWB-BS line bundle, $c_{1}^{a}$ determine the
irreps. On the other hand, the irreps. are given by the maximal
weight $\La$  of the representation. The latter can be written
as a linear combination of the so called {\em fundamental weights},
\cite[\S 14.1]{fh}, with non-negative integer coefficients. Let us
denote these by $\La_{b}$, $b=1,\cdots ,l$. Then,
	\be
	\La =\sum_{b=1}^{l}k_{b}\La_{b}\; ,\;\;\; k_{b}\in \Z^{+}\cup
	\{ 0\}\; .
	\label{q54}
	\ee
This means that to determine the $k_{b}$'s and hence the irrep.,
one needs precisely $l$ ``independent'' first Chern numbers. These
are obtained by integrating (\ref{q53}) over the 2-cells of $G/T$.
The 2-cells are $l$ copies of $S^{2}$ which correspond to the canonical
$SU(2)$ subgroups of $G$. These are generated by the triplets
of the generators $(E_{\a},E_{-\a},H_{\a})$, where $\a$'s are
the $l$ simple roots of ${\cal G}$, and $E_{\a}$ and $H_{\a}$ are
as in eq.\ (\ref{q31}). Denoting these $SU(2)$ subgroups and their
maximal tori by $G_{a}$ and $T_{a}$, respectively,
the 2-cells are given by
	\be
	\s_{a}:=G_{a}/T_{a} = SU(2)/U(1) = S^{2}\; .
	\label{q55}
	\ee
The restriction of the curvature 2-form $\Omega$ on $\s_{a}$ yields
Berry's curvature 2-form, \cite{berry}. Integrating these 2-forms
on $\s_{a}$ gives rise to $l$ identities of the form (\ref{q20}).
Incidentally, in view of the relevance of the system of eq.\ (\ref{q15})
to magnetic monopoles, \cite{monopole}, (\ref{q25})
corresponds to a generalized magnetic monopole whose charge
has a vectorial character with integer components. I shall
return to the discussion of monopoles in section 6.

\section{Berry's Connection and the Riemannian geometry of the
Parameter Manifold}
One of the rather interesting facts about the geometric phase
is that the AA connection ${\cal A}$ is related to the Fubini-Study
metric on the projective space $\cp^{N}$, \cite{page}. In the language
of fibre bundles, the Riemannian geometry of a manifold $X$ means the
geometry of its tangent bundle $TX$. In particular, the Riemannian
metric (the Levi Civita connection) is a metric (resp. a
connection) on $TX$. The statement that the AA connection
is related to the Riemannian geometry of $\cp^{N}$ is equivalent
to say that the universal (AA) bundle
	\[ E:\C\rar E\rar \cp^{N} \]
is related to the tangent bundle
	\[ T\cp^{N}:\C^{N}\rar T\cp^{N}\rar\cp^{N} \; .\]
This is easy to show topologically. The precise relation is demonstrated
in the form of the following identity:
	\be
	Det\,\ll T\cp^{N}\rr = E^{*}\otimes E^{*} \; ,
	\label{q56}
	\ee
where $Det$ means the determinant bundle:
	\[ Det\,\ll T\cp^{N}\rr := \underbrace{
	T\cp^{N}\wedge\cdots\wedge T\cp^{N}}_{N\mbox{-times}} \; , \]
$\wedge$ stands for the wedge product of the vector bundles,
$E^{*}$  is the dual line bundle to $E$, and $\otimes$ is the
tensor product, \cite{egh}. To see the validity of eq.\ (\ref{q56}), it is
sufficient to examine the first Chern classes of both sides. In fact,
since $\cp^{N}$ has a single 2-cell, namely $\cp^{1}=S^{2}$, one can
simply compare the first Chern numbers. It is well-known, \cite{nakahara},
that
	\be
	c_{1}(E)=-1 \; .
	\label{q57}
	\ee
Furthermore, for any vector bundle $V$
	\be
	\hat{c}_{1}\ll Det\, V\rr =\hat{c}_{1}\ll V\rr \; .
	\label{q58}
	\ee
Also it is not difficult to show that
	\be
	c_{1}(T\cp^{N} )=c_{1}(T\cp^{1})=\chi (S^{2})=2\; ,
	\label{q59}
	\ee
where $\chi$ stands for the Euler-Poincar\'{e} characteristic.
Eqs.\ (\ref{q58}) and (\ref{q59}) imply that
	\[ c_{1}\ll Det\, T\cp^{N}\rr = 2 \; .\]
The last equality together with the fact that
	\[ c_{1}(E^{*})=-c_{1}(E) \]
and eq.\ (\ref{q57}) are sufficient to establish the validity of
eq.\ (\ref{q56}).

The existence of this relationship between the AA connection and the
Riemannian metric on $\cp^{N}$ has triggered  the investigation of
a similar pattern in the BS approach, \cite{poland}. In \cite{poland},
the authors discuss the case of a general Hamiltonian with a dynamical
group $G$ and a parameter space $G/H$ where $H$ is a closed subgroup
of symmetries of the Hamiltonain. The analysis presented above
seems to include all these cases. In the following section, I will show that
the system of eq.\ (\ref{q25}) has a universal character. In other
words, all the cases discussed in \cite{poland} can be reduced to the
one given by (\ref{q25}).
In all these cases the parameter space, $G/H$, is a submanifold
of $FU(m):=U(m)/T^{m}$, $T^{m}:=\ll U(1)\rr^{m}$, which is itself embedded
into $\cp^{\infty}$. Hence, the results of \cite{poland} are expected because
\begin{itemize}
\item the BS bundle (connection) is the pullback (restriction) of the universal
bundle $E$;
\item $E$ is related to $T\cp^{N}$, via eq.\ (\ref{q56}).
\end{itemize}

\section{Reduction of the Non-Adiabatic Phase to the
Adiabatic Phase for the Cranked Hamiltonians}
Let us consider an arbitrary $m\times m$ Hamiltonian $H$ acting
on ${\cal H}=\C^{m}$. $H$ can be viewed as an element of the
(real) vector space of all complex $m\times m$ dimensional
hermitain matrices. It is very easy to compute the real dimension
of this space and find out that it is equal to $m^{2}$. Thus,
$H$ can be written as a linear combination of $m^{2}$ linearly
independent hermitain matrices. Incidentally, the generators
$J_{i}$ of $U(m)$ form a set of $m^{2}$ such matrices. This
simply indicates that one can always express $H$ in the form of
eq.\ (\ref{q25}). This may be seen as a realization of the Peter-Weyl
theorem, \cite{segal}. The particular representation of $H$ given
by eq.\ (\ref{q25}) with $G=U(m)$ for some $m\!\in\!\!\Z^{+}$ might not
be a practical choice. For example, the quadratic Hamiltonian
	\[ H=\sum_{i,j=1}^{3}Q_{ij}\, \s_{i}\otimes \s_{j}\; , \]
with $\s_{i}$ being Pauli matrices, \cite{poland,babuch},
is more managable in this form than in the form of eq.\ (\ref{q25})
with $J_{i}$ chosen to be the generators of $U(4)$. However, in
principle one can always use the linear representation, eq.\ (\ref{q25}).

Actually, one can use the generators of $SU(m)$ rather than $U(m)$.
This is emphasized in \cite{as}. It can be directly justified  by recalling
that the $(m^{2}-1)$ generators of $SU(m)$ are also linearly
independent and these together with the $(m\times m)$ identity
matrix $I$ provide a basis for the space of $(m\times m)$ hermitian
matrices. The Hamiltonian $H$ can then be written as a linear
combination in this basis. Clearly, the term proportional to $I$
does not contribute to the geometric phase. This is often used
as an indication of the geometric nature of Berry's phase, \cite{bohm}.

An advantage of the linear representation is that it allows one to
use the knowledge about the universal bundles and BWB theorem
directly. In particular, in some cases,
it is possible to obtain the non-adiabatic analog of the BS line
bundle and the connection $A$. The first example of this is
presented in \cite{mb}. In this section, I will show that since
the above argument does not refer to the adiabaticity of the system,
one can always reduce the Hamiltonian to the linear form. Moreover,
if the time dependence of the corresponding
linear Hamiltonian is realized by cranking of the  initial
Hamiltonian along a fixed direction \cite{wang}, then one can
obtain a non-adiabatic analog $\tilde{A}$ of
Berry's connection $A$ as a pullback connection
one-form. The geometric phase is then identified with the associated
holonomy of the loops in the space of parameters. This
is quite remarkable because it means that one does not need to
solve the Schr\"{o}dinger equation, provided that the function $F$ that
induces $\tilde{A}$ as a pullback one-form is given.
Wang \cite{wang} has presented a procedure that essentially computes
$F$. Nevertheless, he does not even label this function nor does he
implement the idea of universal  bundles. Let us see how the conditions
introduced in \cite{mb} are realized in for cranked Hamiltonians.
These conditions are:
\begin{enumerate}
\item The cyclic states are the eigenstates of a hermitian operator
$\tilde{H}$ which depends parametrically on the points of the
parameter manifold $M$, i.e., the cyclic states are eigenstates
of $\tilde{H}(\xo )$ with $\xo =x(t=0)$.
\item $\tilde{H}$ is related to the Hamiltonian according to
	\be
	\tilde{H}(x)=H(F(x))=(HoF)(x)\; ,
	\label{q60}
	\ee
where $F:M\to M$ is some smooth function, such that in the adiabatic
limit, $F$ approaches to the identity map.
\end{enumerate}
Let us first see how the first condition is fulfilled for any periodic
Hamiltonian. According to a result of Floquet theory, \cite{moore}, the time
evolution operator for any periodic Hamiltonian is of the form
	\be
	{\cal U}(t)=Z(t)\, e^{it\tilde{H}}\; ,
	\label{q61}
	\ee
where $\tilde{H}$ is a time independent hermitian operator and
$Z$ is a periodic unitary operator with the same period as the
Hamiltonian, i.e.,
	\bea
	Z(t+T)&=&Z(t)\; , \label{q62} \\
	Z(0)&=&1\; . \nn
	\eea
A simple proof of this statement, i.e.,
eq.\ (\ref{q61}), is presented in the appendix.
Clearly, one has
	\be
	{\cal U}(T)=e^{iT\tilde{H}} \; ,
	\label{q63}
	\ee
which justifies the first condition. The second condition can
be seen to hold for the cranked Hamiltonians
either by refering to the work of Wang \cite{wang}
or following the argument used in the discussion of the
transformation of the Hamiltonian into the linear form.
The latter is quite straightforward. One simply starts by realizing that since
$\tilde{H}$ is hermitian, it can also be written in the linear
form:
	\be
	\tilde{H}(x_{0})=\sum_{i=1}^{d} \tilde{x_{0}}^{i}\, J_{i}\; ,
	\label{q25'}
	\ee
where $\tilde{\xo}:= \lll \tilde{\xo}^{i}\rrr\in M$ must depend on the
Hamiltonian (\ref{q25}),
and consequently on $C \subset M$. However, for the cranked Hamiltonians
the time dependence of the Hamiltonian is governed by the action of
a one parameter subgroup of $G$, i.e., the operator $U(t)$ of eq.\
(\ref{q27}) is given by
	\[ U(t):= \exp \ll i\,\omega t\, n_{\a}E_{\a}\rr\;\;\mbox{with}\;\;
	n_{\a}=\mbox{const.}\; ,\]
where $\omega$ and  $(n_{\a})$ are called the cranking rate and
direction, respectively. It is
clear that for such systems $\tilde{x_{0}}$ can only depend on
the initial Hamiltonian and thus on $\xo$.
The function $F$ is defined by
	\be
	\tilde{\xo}=: F(\xo) \; .
	\label{q63.1}
	\ee
The only problem is that in some cases, depending on the value
of the slowness parameter $\n$ ($\omega$), $F$ may be discontinous
or even multi-valued.
This happens in the case of eq.\ ~(\ref{q15}) for $\n=\omega /b =1$.
But in the generic
case $F$ is smooth and the second condition holds as well.
The non-adiabatic analog of the BS line bundle is then given by
	\be
	\tilde{L}:=F^{*}(L)\; .
	\label{q64}
	\ee
It is endowed with the non-adiabatic connection one-form
	\be
	\tilde{A}:=F^{*}(A)\; .
	\label{q65}
	\ee
For completeness, let me briefly review the arguments of \cite{mb}
which lead to eqs.\ (\ref{q64}) and (\ref{q65}). The basic idea is
that the existence of $\tilde{H}$ which satisfies eq.\ (\ref{q25'})
allows one to imitate Berry's treatment of the adiabatic systems.
The energy eigenstate vectors $|n,x\kt$ are replaced by the eigenstate
vectors $|\tilde{n},x\kt$ of $\tilde{H}(x)$. In view of eq.\ (\ref{q60}),
these are given by
	\be
	|\tilde{n},x\kt =|n,\tilde{x}\kt =|n,F(x)\kt \; .
	\label{q66}
	\ee
The non-adiabatic line bundle $\tilde{L}$ is obtained from the universal
line bundle $E$ via the non-adiabatic analog of the map $f$ of eq.\
(\ref{q12}).
Denoting the latter by $\tilde{f}:M\to \ph$, one has
	\bea
	\tilde{f}(x)&:=&|\tilde{n},x\kt\br\tilde{n},x|\nn \\
	&=& |n,F(x)\kt\br n,F(x)|\nn \\
	&=& (foF)(x) \; . \nn
	\eea
Then, using the functorial property of the pullback operation one
shows that
	\bea
	\tilde{L}&=&\tilde{f}^{*}(E)\nn \\
	&=&(foF)^{*}(E)\nn \\
	&=&(F^{*}of^{*})(E) \nn \\
	&=& F^{*}(L) \; ,\label{q67}
	\eea
where in the last equality eq.\ (\ref{q12.1}) is used. This proves
eq.\ (\ref{q64}). The proof of eq.\ (\ref{q65}) is identical. An
important observation is that unlike $|n,\xo\kt$ the initial state vectors
$|\tilde{n},\xo\kt$ undergo  exact cyclic evolutions.
\section{More on Parameter Spaces and Monopoles}
In the discussion of the the relation between the BS connection and the
Riemannian structure on the parameter space, the parameter space is
taken to be $M=G/H$, for some arbitrary closed subgroup $H$ of $G$,
\cite{poland}. It can be shown that all these cases are
included in the analysis of the linear system eq.\ (\ref{q25}).

In section 3, I argued that depending on the (maximal weight $\La$ of the)
irrep. of $G$, $M$ is of the form $G_{\C}/P\subset G/T$, where $P$ is defined
by
eq.\ (\ref{q46.1}). Let us consider the Weyl chamber ${\cal W}$ of
$\Upsilon^{*}$ with respect to which the positive and the negative
roots are distinguished, \cite{fh}. If $\La$ happens to lie on at
least one of the walls of ${\cal W}$ then $B$ is a proper subgroup of
$P$, otherwise $P=B$.  The universal character of the linear Hamiltonian
is also realized in that all the homogeneous spaces of $G$ can be
obtained as $G_{\C}/P$ by choosing $\La$ appropriately. In fact, this is
the basic idea of the classification of the compact homogeneous spaces of
semisimple Lie groups. Therefore, in principle one should be able
to reproduce the results of \cite{poland} using the relation of
Berry's phase to the theory of universal bundles.

Let us consider the group $G=SU(3)$ in its defining (standard)
representation. $SU(3)$ is of rank $l=2$. So any irrep. is given by
two integers. The standard representation is itself a
fundamental representation, namely $(k_{1}=1 ,k_{2}=0)$,
\cite{fh}. The maximal weight is on a wall of ${\cal W}$
and the Borel subgroup of upper triangular matrices
in $SL(3,\C )=SU(3)_{\C}$ is a proper subgroup of $P$. The
subgroup $P$ of  $SL(3,\C )$ consists of the elements of the form:
	\[ \ll \begin{array}{ccc}
	* & * & * \\
	* & * & * \\
	0 & 0 & * \end{array} \rr\; ,\]
where $*$ are complex numbers, \cite{fh}.
The parameter space is $M=SL(3,\C )/P=SU(3)/U(2)=\cp^{2}=\ph$.
It is interesting to see that in this case the parameter space
$M$ and projective Hilbert space $\ph$ are identical. In fact,
this is true for all $SU(N+1)$ groups. The defining representation
corresponds to $(k_{1}=1,k_{2}=\cdots =k_{N}=0)$ and the
parameter space is $M=SU(N+1)/U(N)=\cp^{N}=\ph$. Therefore the
inducing map $f$ maps $\cp^{N}$ to itself for all $N>1$.

The situation is different for the octet representation of $SU(3)$.
In this case one has $k_{1}=k_{2}=1$. $\Lambda$ lies in the interior
of ${\cal W}$, $P=B$, and the parameter space is the full flag manifold
$M=SU(3)/U(1)\times U(1)$. The map $f$ maps  $M$ into $\ph =\cp^{7}$.
\footnote{Note that this representation is 8-dimensional, i.e., the
representation space for $SL(3,\C )$ is $\C^{8}$. Hence ${\cal H}=\C^{8}$.}

For $G=SU(2)$, it is well-known that the system of eq.\ (\ref{q15})
is related to the magnetic monopoles, \cite{monopole}. The relation
of monopoles to the gauge theories and their generalization to
arbitrary compact semisiple gauge groups have been studied in the late
seventies, \cite{goddard}. These generalized monopoles are called
{\em non-abelian} or {\em multi-monopoles} for general groups and
{\em color monopoles} for $SU(3)$, \cite{cho}. They are topologically
classified by an associated set of $l$ integers where $l$ is the
rank. These are called the {\em topological charges} of the monopole
and they are defined as elements of the second homotopy group
$\pi_{2}(G/H)$, where $H$ is the group of the symmetries of
a ground state of the Higgs fields (a minimum of Higgs potential),
\cite{goddard}. For $G=SU(3)$, there are two possibilities. Either,
	\[\mbox{I)}\; H=U(2) \;\;\;\mbox{or}\;\;\;\mbox{I\hspace{-.4mm}I)}
	\; H=T=U(1)\times U(1)\; .\]
These cases have been studied in almost every article written in
this subject, e.g. see \cite{perelomov}, \cite{goddard} and
references therein.

If $G$ is simply connected then a result
of algebraic toplogy indicates that
	\[\pi_{2}(G/H) =\pi_{1}(H)\; .\]
Applying this result to $G=SU(3)$ one finds
	\bea
	\mbox{I)}&&\pi_{2}(SU(3)/U(2))=\pi_{1}(U(2))= \Z\nn \\
	\mbox{I\hspace{-.4mm}I)}&& \pi_{2}(SU(3)/U(1)\times U(1))=
	\pi_{1}(U(1)\times U(1))= \Z\oplus\Z \; .\nn
	\eea
Thus, for I) and I\hspace{-.4mm}I) one has, respectively, one and two
topological charges. This is precisely the case with the
topological charges of the geometric phase defined earlier. The
same correspondence holds for arbitrary compact, connected
semisimple Lie groups.

The possible relevance of the topological charges of monopoles
to the reperesentations of the group have been conjectured
by Goddard, {\em et al}.\ , \cite{gno}. Although, the analysis of the present
paper does not prove their conjecture, it provides a formula for
the topological charges as integrals of the first Chern class,
defined by Berry's connection, over the 2-cells $\s_{a}$ of section
3. There is a simple topological explanation for the correspondence
of the topological charges of the monopoles and those of the geometric phase.
This can be summarized in the identity
	\[ \pi_{2}(G/H)=H_{2}(G/H ,\!\Z ) \; ,\]
where $H_{2}(.,\!\Z )$ denotes the second homology group. This
identity is a consequence of {\em Hurewicz theorem}, \cite{at},
where one uses the fact that $\pi_{1}(G/H)=H_{1}(G/H)=0$.
The 2-cells $\s_{a}$ are indeed the generators of $H_{2}(G/T,\!\Z)$.
For $H\neq T$ some of them may be smashed to a point as is the
case for $G=SU(3)$ and $H=U(2)$.

\section{Conclusion}
The relevance of the phenomenon of Berry's phase to Borel-Weil-Bott
theorem and specially to the theory of universal bundles
is appealing not only for the aesthetical reasons but
also for its allowing for a better understanding of the non-adiabatic phase.
Moreover, it sheds light on a number of issues such as the determination of the
appropriate parameter space and the relation between the geometry
of the parameter space and the geometric structure of the phase.
The BWB theorem leads to the definition of of a set of topological
charges which determine the topology of the BS line bundles. These
seem to be related, if not identical, to the topological charges
of non-abelian monopoles. The integral nature of these charges
is a consequence of the topological properties of the first Chern
class. The latter is essentially the reason for the quantization
of the charges of the monopoles.

\section*{Acknowledgements}
The idea of the relation between Berry's phase and the universal bundles
was originated in a project I did in the course of Methods of Mathematical
Physics given by Prof.\ Arno Bohm in the summer of 1991. The key
idea was to recognize the utility of the projective space $\cp^{\infty}$
in both subjects. I had learned the corresponding algebraic topology
from Prof.\ Gary Hamrick.
I would also like to acknowledge Prof.\ Luis J.\ Boya for introducing me to
the Borel-Weil-Bott theorem and his invaluable suggestions,
Prof.\ Pierre Cartier for teaching me
a proof of eq.\ (\ref{q61}), Prof.\ David Auckley for many fruitful
discussions, Prof.\ Larry Biedenharn for several constructive comments,
and my supervisor Prof.\ Bryce DeWitt who has
encouraged me to work in this subject, although it was not related
to the subject of my dissertation.
\section*{Appendix: A note on Floquet theory}
The following is a proof which I learned its main idea from
Prof.\ Pierre Cartier.

Let $H=H(t)$ be a T-periodic selfadjoint operator serving as the
nonconserved Hamiltonian of a quantum system, i.e.,
	\[ H(t+T)=H(t)\;\;\; ,\;\;\;\forall t\in [0,T] \; .\]
Let ${\cal U}(t)$ be the time evolution operator which satisfies
the Schr\"{o}dinger equation,
	\bea
	\frac{d}{dt}\, {\cal U}(t)&=&-i\, H(t)\, {\cal U}(t)\label{a1} \\
	{\cal U}(0)&=&1\; .\nn
	\eea

{\bf Theorem}: There exist a time independent selfadjoint operator
$\tilde{H}$ and a T-periodic unitary operator $Z=Z(t)$ such that
${\cal U}(t)$ is of the form
	\be
	{\cal U}(t)=Z(t)\, e^{it\, \tilde{H}} \; ,
	\label{star}
	\ee
and
	\[ Z(0)=Z(T)=1\; .\]

{\bf Proof}: Let $V(t):={\cal U}(t+T)$, then $V$ satisfies the following
schr\"{o}dinger equation:
	\bea
	\frac{d}{dt}V(t)&=&-i\, H(t)\, V(t) \label{a2} \\
	V(0)&=&{\cal U}(T)=:C\nn\; .
	\eea
The operator $V'(t):={\cal U}(t)\, C$ satisfies eq.\ (\ref{a2})
as well. Then the uniqueness of the solution of this differential
equation implies that $V(t)=V'(t)$, i.e.,
	\be
	{\cal U}(t+T)={\cal U}(t)\, {\cal U}(T)\; .
	\label{a3}
	\ee
One can easily show that (\ref{star}) satisfies eq.\ (\ref{a3}).
Thus it is the unique solution. In fact, it is not difficult to
construct a pair ($Z(t),\tilde{H}$) which satisfies eq.\ ~(\ref{star}).

Let $t=nT+t_{0}$ for some $n\!\in\!\!\Z$ and $t_{0}<T$. $n$ and $t_{0}$ are
uniquely determined for given $t$.
Applying eq.\ ~(\ref{a3}) repeatedly, one has
	\[ {\cal U}(t)={\cal U}(t_{0})\ll {\cal U}(T)\rr^{n}=
	{\cal U}(t_{0})\, C^{n}\; .\]
Relabelling $C$ by $e^{i\tilde{H}'}$, and noting $n=\frac{t-t_{0}}{T}$,
one obtains
	\bea
	{\cal U}(t)&=&{\cal U}(t_{0}).e^{\frac{i(t-t_{0})}{T}
	\tilde{H}'} \nn \\
	&=& {\cal U}(t_{0}).e^{-\frac{it_{0}}{T}\tilde{H}'}.
	e^{\frac{it}{T}\tilde{H}'} \nn \\
	&=&Z(t)\, e^{it\,\tilde{H}} \; ,\nn
	\eea
where
	\[ Z(t):={\cal U}(t_{0})\, e^{-\frac{it_{0}}{T}\tilde{H}'}\; ,\]
and
	\[ \tilde{H}:=\frac{\tilde{H}'}{T} \; .\]
Clearly, $\tilde{H}$ is selfadjoint and $Z(t)$ is unitary.
Furthermore, $Z(t)$ satisfies
	\bea
	Z(t+T)&=&Z(t)\nn \\
	Z(0)&=&{\cal U}(0)\: =\: 1 \; ,\nn
	\eea
by construction. One must however note that $\tilde{H}'$ is not
unique, nor is the decomposition (\ref{star}).

\end{document}